\documentclass[pra,preprint,showpacs]{revtex4}
\usepackage{graphicx}
\usepackage{amssymb}
\usepackage{subfigure}
\usepackage{color}
\newcommand{\cor}{ }

\begin{document}

\title{Collapse of ultrashort spatiotemporal pulses described by the cubic generalized Kadomtsev-Petviashvili equation }
\author{Herv{\'e} Leblond$^{1}$, David Kremer$^{1}$, and  Dumitru Mihalache$^{2,3}$ }
\affiliation{
$^{1}$Laboratoire de Photonique d'Angers,
 Universit\'e d'Angers, 2 Bd. Lavoisier, 49045 Angers Cedex 01, France\\
$^{2}$Horia Hulubei National Institute for Physics and Nuclear Engineering (IFIN-HH),
407 Atomistilor, Magurele-Bucharest, 077125, Romania\\
$^{3}$Academy of Romanian Scientists, 54 Splaiul Independentei, Bucharest 050094, Romania}

\begin{abstract}

By using a reductive perturbation method, we derive from Maxwell-Bloch equations, a cubic generalized Kadomtsev-Petviashvili equation for ultrashort spatiotemporal optical pulse propagation in cubic (Kerr-like) media, without the use of the slowly varying envelope approximation.
We calculate the collapse threshold for the propagation of few-cycle spatiotemporal pulses described
by the generic cubic generalized Kadomtsev-Petviashvili equation by a direct numerical method, and compare it to analytic results based on a rigorous virial theorem. Besides, typical evolution of the spectrum (integrated over the tranverse spatial coordinate) 
is given and a strongly asymmetric spectral broadening of ultrashort spatiotemporal
pulses during collapse is evidenced.

\end{abstract}


\maketitle

\section{Introduction}

The rapid progress over the past two decades in the area of ultrafast optics has led to the production, manipulation and control of pulses with durations down to a few optical cycles (see, e.g., the comprehensive
review \cite{review}). The theoretical and experimental studies of few-cycle pulses (FCPs) have
opened the door to a series of applications in various fields
such as light matter interaction, high-order harmonic generation,
extreme nonlinear optics \cite{Wegener}, and attosecond physics \cite{atto1,atto2}.
On the theoretical arena three classes of main dynamical models for FCPs have been put forward: (i) the quantum approach \cite{tan08,ros07a,ros08a,naz06a}, (ii) the refinements within the framework of the slowly varying envelope approximation (SVEA) of the nonlinear Schr\"{o}dinger-type envelope equations \cite{Brabek_PRL,Tognetti,Voronin,Kumar}, and non-SVEA models \cite{quasiad,leb03,igor_jstqe,igor_hl,interaction}.  
In media with cubic (Kerr-type) optical nonlinearity the physics of (1+1)-dimensional FCPs  can be adequately described beyond the SVEA by using different dynamical models, such as the modified Korteweg-de Vries (mKdV) \cite{quasiad}, sine-Gordon (sG) \cite{leb03,igor_jstqe}, or mKdV-sG equations \cite{igor_hl,interaction,LM_PRA_2009}.
It is worthy to notice that the physics of the (1+1)-dimensional FCPs is well described by the generic mKdV-sG equation \cite{igor_hl}; this quite general model was also derived and studied in Refs. ~\cite{saz01b,bugay}.

Another class of conceptually important
optical problems for which the SVEA approach does
not apply in the femtosecond regime includes multidimensional
spatiotemporal optical solitons (alias ``light bullets"), formed
by the competing diffraction, dispersion, and
quadratic \cite{LB_quadratic} or cubic \cite{LB_cubic} nonlinearity.
Thus for the adequate description of multidimensional FCPs, a non-integrable generalized Kadomtsev-Petviashvili (KP) equation \cite{KP} (a two-dimensional version of the mKdV model) was introduced for (2+1)-dimensional few-optical-cycle spatiotemporal soliton propagation in {\it cubic} nonlinear media beyond the SVEA~\cite{igor2000,igor_matcom}. Recently, by using a powerful reductive perturbation technique \cite{tutorial}, or a multiscale analysis, a generic KP evolution equation governing the propagation of femtosecond spatiotemporal optical solitons in {\it quadratic} nonlinear media beyond the SVEA was also put forward \cite{kpopt}. 

Transparency of the medium is obviously required by soliton propagation. 
It implies that all transition frequencies of the medium are far from the typical frequency of the wave.
They can be either well above or well below the latter.
 As shown in (1+1) dimensions \cite{leb03}, a wave frequency much lower than the resonance frequency corresponds to a long wave approximation and to a mKdV model,
 while a wave frequency much higher than it corresponds  to a short wave approximation and a sG model.
Later, a generic mKdV-sG model has been derived for the case of two transition lines, one well above, and the other one well below the wave frequency \cite{igor_hl},
and is the most general non-SVEA model proposed to describe FCP propagation in Kerr media \cite{LM_PRA_2009}. It is expected to remain valid 
in the general case, where two sets of resonance lines are present instead of two single transitions. 
Our ultimate goal is to generalize the generic mKdV-sG model to (2+1) dimensions.
However, both long- and short wave approximations for a simple two-level model have not yet been rigorously derived.
It might be relevant to consider the full set of atomic levels, and it would be 
necessary in order to compute the nonlinear coefficients in a quantitative way. However, as a preliminary approach and for the sake of tractability, we need to consider a simplified model, but a natural question arises, why we consider a two-level model and not, e.g., a four-level one? As 
written above, it has been previously shown \cite{leb03,igor_hl} that if we consider only two separate transitions, one with resonance frequency below the optical range, and the second one above it, 
the FCP propagation is considerably affected. The most relevant refinement of the two-level model is 
thus a $2\times 2$ level one.
Our preliminary computations for the model of two sets of two levels show that in the long wave approximation considered in the present work,
both the dispersive and the nonlinear coefficients merely sum up. In fact, we expect that the model equation obtained 
for the complete model with arbitrary number of levels, when all resonance lines are well above the wave frequency, has the same form as the one derived for the two-level model, only with modified values of nonlinear coefficients. 
However, this is not the case when the resonance line is well below the wave frequency, since the pulse evolution involves the difference of populations between the levels in a nonlinear way.
In this situation, we expect a more complicated evolution equation for FCPs when more than two levels are involved. 
This general model, which might be, in our opinion, the most relevant correction to the two-level model, is left for further investigation.
On the other hand, the two-level system has many interesting features by itself: It is in some sense equivalent 
to a classical oscillator, as can be shown from the computation of the nonlinear susceptibilities \cite{boyd}, 
or from the derivation of the 
KdV model for FCPs propagation in quadratic nonlinear media \cite{kdvopt}. 
Moreover, estimations of the value of the nonlinear coefficient $n_2$ drawn from the classical model have shown a very good agreement with experimental data, 
as, e.g., in the case of the model developed in Ref. \cite{bgo}.

Thus there are three main issues which deserve to be investigated: the first one, which constitutes the aim of the present manuscript, is the 
long-wave approximation leading to the cubic generalized Kadomtsev-Petviashvili (CGKP) equation, which yields 
collapse. The second and the third worth studying open problems are, respectively, the short wave approximation and the investigation of the full model with both mKdV and sG terms.
On this ground, we consider that the study of the CGKP model, including the derivation of the CGKP equation from a two-level model, is an important step in the adequate description of (2+1)-dimensional FCPs.

The aim of this paper is to derive and to study a cubic generalized Kadomtsev-Petviashvili partial differential
equation, which describes the dynamics of
(2+1)-dimensional spatiotemporal pulses in cubic nonlinear
media beyond the SVEA model equations,
starting from the Maxwell-Bloch equations for a set of two-level atoms. This paper is organized as follows. 
In Sec. II we derive the generic CGKP equation and we study by adequate numerical methods the propagation dynamics,
the nonlinear difraction  and,
in the case of anomalous dispersion,
the collapse of ultrashort spatiotemporal pulses.  We calculate in Sec. III the collapse threshold for the propagation of ultrashort spatiotemporal pulses described
by the CGKP equation by both a direct numerical method and by 
an analytical method, which is based on a rigorous virial theorem.   In Sec. IV, the evolution of the spectrum (integrated over the transverse coordinate) 
is given and a strongly asymmetric spectral broadening of ultrashort
pulses during collapse is put into evidence. Section V presents our conclusions.

\section{The cubic generalized Kadomtsev-Petviashvili equation and its numerical computation}

For a Kerr (cubic) medium a CGKP equation can be derived from Maxwell-Bloch equations using the powerful reductive perturbation method \cite{tutorial}.
\cor{
As said in the introduction, we consider a set of two-level atoms with the Hamiltonian
\begin{equation}
H_0=\hbar\left(\begin{array}{cc}\omega_a&0\\0&\omega_b\end{array}\right),\label{hamil}
\end{equation}
where $\Omega=\omega_b-\omega_a>0$ is the frequency of the transition.
The evolution of the electric field $E$ 
 is described by the wave equation
\begin{equation}
\left(\partial_y^2+\partial_z^2\right)E=\frac1{c^2}\partial^2_t\left(E+4\pi P\right),\label{max}
\end{equation}
where $P$ is the polarization density.
 The light propagation is
 coupled with  the medium by means of a dipolar electric momentum \begin{equation}
\mu=\left(\begin{array}{cc}0&\mu\\\mu^\ast&0\end{array}\right)
\end{equation} directed along the same direction
 $x$ as the electric field, according to
 \begin{equation}
 H=H_0-\mu E,\label{eq17}
 \end{equation}
 and the polarization density $P$ along the $x$-direction is
 \begin{equation}
 P=N \mathrm{Tr}\left(\rho\mu\right),\label{eq18}
 \end{equation}
 where $N$ is the volume density of atoms and $\rho$ the density matrix.
Since, as shown in \cite{leb03}, the relaxation  can be neglected, the density-matrix evolution equation (Schr{\"o}dinger equation) reduces to
 \begin{equation}
 i\hbar\partial_t\rho=\left[H,\rho\right].\label{schr}\label{eq19}
 \end{equation}
}

\cor{
Transparency implies that the characteristic frequency $\omega_w$ of the considered
radiation (in the optical range)  strongly differs from  the resonance frequency $\Omega$ of the atoms.
Here, as explained in the introduction, we assume that $\omega_w$ is much smaller than $\Omega$.
This motivates the introduction of the slow variables
\begin{equation}
\tau=\varepsilon\left(t-\frac zV\right),\quad
\zeta=\varepsilon^3z,\quad
\eta=\varepsilon^2 y,\label{scal}
\end{equation}
$\varepsilon$ being a small parameter.
The delayed time $\tau$ involves propagation at some speed $V$ to be determined.
It is assumed to vary slowly in time according to the assumption $\omega_w\ll \Omega$.
The pulse shape described by the variable $\tau$ evolves even more slowly in time, the
corresponding scale being that of variable $\zeta$.
The transverse spatial variable $y$ has an intermediate scale as usual in KP-type expansions \cite{tutorial}.
}

\cor{
 Next we use the reductive perturbation method as developed in Ref. \cite{tutorial}. To this aim we expand the electric field $E$ as power series of a small parameter $\varepsilon$:
\begin{equation}
E=\varepsilon E_1+\varepsilon^2E_2+\varepsilon^3E_3+\ldots,\label{exp}
\end{equation}
as in the standard mKdV-type expansions~\cite{tutorial}.
A weak amplitude assumption is needed in order that the nonlinear effects arise at the same propagation distance scale
as the dispersion does, which justifies that the expansion of $E$ begins at order $\varepsilon$.
The polarization density $P$ is expanded in the same way.
}

\cor{The expansion [Eqs. (\ref{scal}) and (\ref{exp})] is then reported into the basic 
 equations [Eqs. (\ref{hamil}-\ref{eq19})], and solved order by order.
The computation is very close to the (1+1)-dimensional case (see Ref. \cite{leb03})
in what concerns nonlinearity and dispersion, while 
the treatment of dispersion and the dependency with respect to the transverse (spatial) variable $\eta$
is fully analogous to the case of quadratic nonlinear media, see Ref. \cite{kpopt}.}
As a result we get the model equation
\cor{
\begin{equation}
\partial_\zeta\partial_\tau E_1=A\partial_\tau^4 E_1+B\partial_\tau^2 \left( E_1\right)^3+\frac V2\partial_\eta^2E_1,\label{mkp}
\end{equation}
}
which is a  CGKP equation.\cor{
The dispersion and nonlinear coefficients $A$ and $B$ in the above (2+1)-dimensional evolution equation are
\begin{equation}
A=\frac{4\pi N|\mu|^2}{nc\hbar \Omega^3},\quad\mbox{and}\quad
B=\frac{8\pi N |\mu|^4}{nc\hbar^3\Omega^3},
\end{equation}
respectively,   the refractive index being
\begin{equation}
n=\sqrt {1+\frac{8\pi N|\mu|^2}{\hbar\Omega}}\;.
\end{equation}
The  CGKP model equation (\ref{mkp}) is generalized by expressing 
the coefficients $A$ and $B$  in terms of dispersion relation $k(\omega)$ and third-order nonlinear susceptibility $\chi^{(3)}$. They can indeed be written as}
\begin{equation}
A=\frac16\left.\frac{d^3k}{d\omega^3}\right\vert_{\omega=0},\label{coefa}
\end{equation}
\begin{equation}
B=-\left.\frac{6\pi}{nc}\chi^{(3)}_{xxxx}(\omega ,\omega ,\omega ,-\omega )\right\vert_{\omega=0}.\label{coefb}
\end{equation}
Also, the coefficient $V$ in Eq. (\ref{mkp}) is the group velocity, $V=\left.\frac{d\omega}{dk}\right\vert_{\omega=0}$.

\cor{
By means of an adequate linear rescaling, the  CGKP equation (\ref{mkp}) can be reduced to the normalized form
\begin{equation}
\left(u_Z+\sigma_1 u^2u_T+\sigma_2 u_{TTT}\right)_T=u_{YY}, \label{mkpnor}
\end{equation}
where $\sigma_1=\mbox{sgn}(-B)$ and $\sigma_2=\mbox{sgn}(-A)$. 
In the frame of the Maxwell-Bloch equations, $\sigma_1=\sigma_2=-1$, hence the nonlinearity and dispersion
yield temporal self-compression, but nonlinearity and diffraction tend to defocuse the FCP.
}

The  CGKP equation (\ref{mkpnor}) is solved by means of the fourth order Runge Kutta exponential time differencing (RK4ETD)
scheme \cite{cox02}. It involves one integration with respect to $\tau$. The inverse derivative is computed by means of a 
Fourier transform, which implies that the integration constant is fixed so that the mean value of the inverse derivative
is zero, but also that the linear term is replaced with zero, i.e.,  the mean value of the function $u_{YY}$ is set to zero.
For low frequencies, the coefficients of the RK4ETD scheme are computed by means of series expansions, to avoid catastrophic consequences of limited numerical accuracy.

We consider in our simulations input data in the form
 \begin{equation}\cor{
u(T,Y,Z=0)=A\; e^{-T^2/p^2-Y^2/q^2}\;\sin (\omega T).\label{data}}
\end{equation}

\begin{figure}\begin{center}
\subfigure[ ]{\includegraphics[width=7cm]{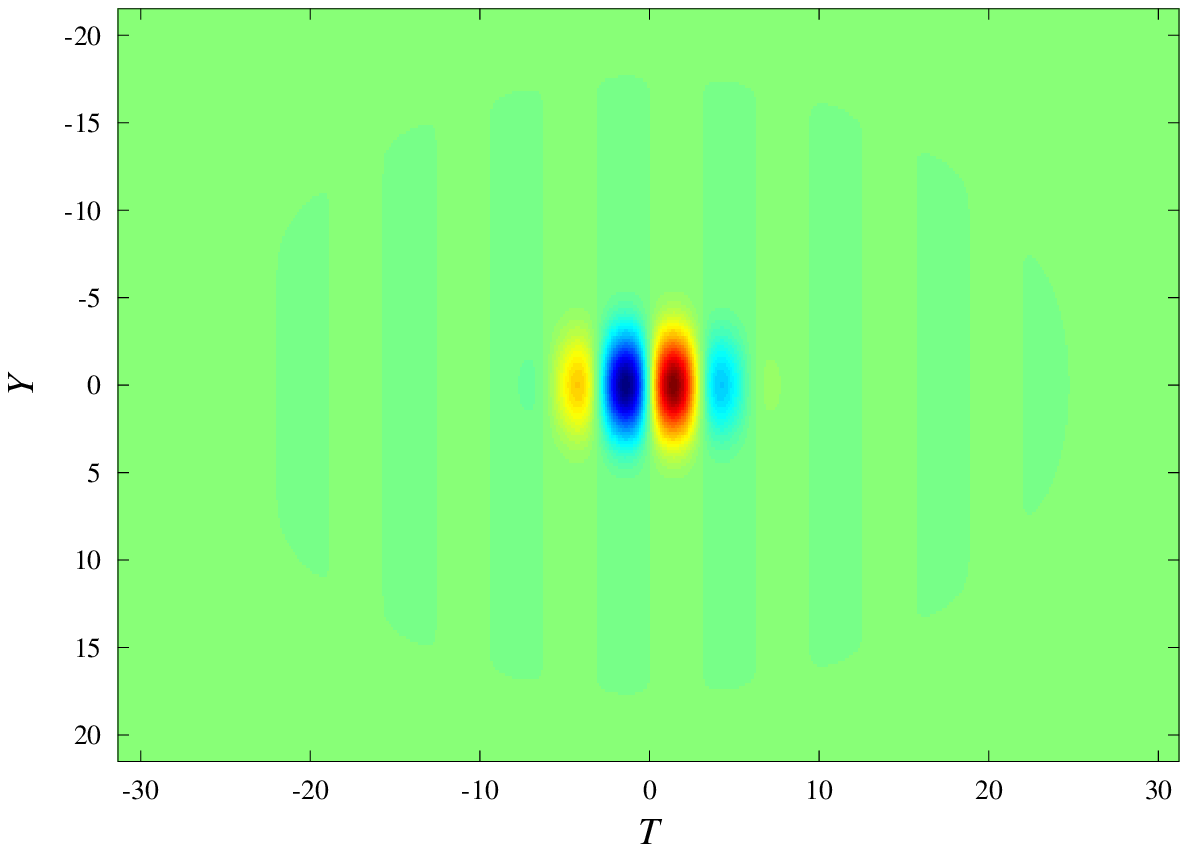}}
\subfigure[ ]{\includegraphics[width=7cm]{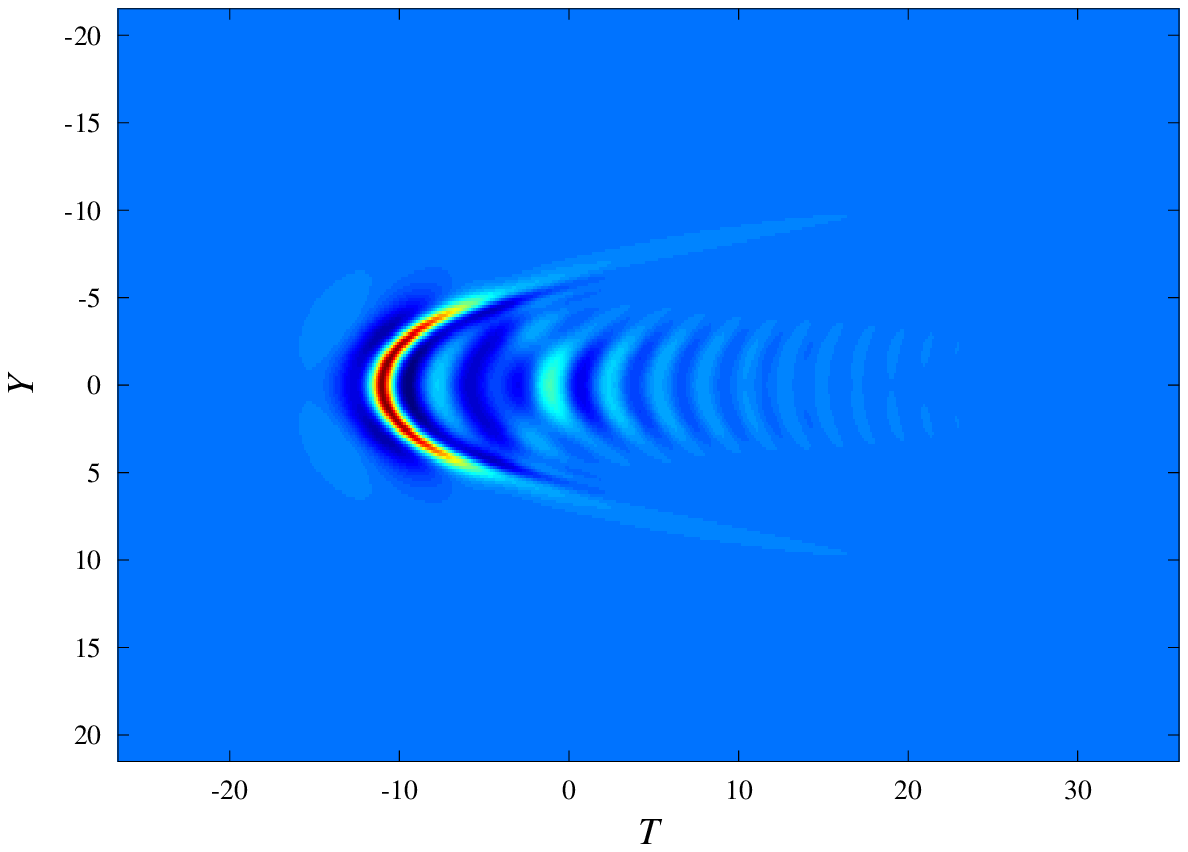}}
\subfigure[ ]{\includegraphics[width=7cm]{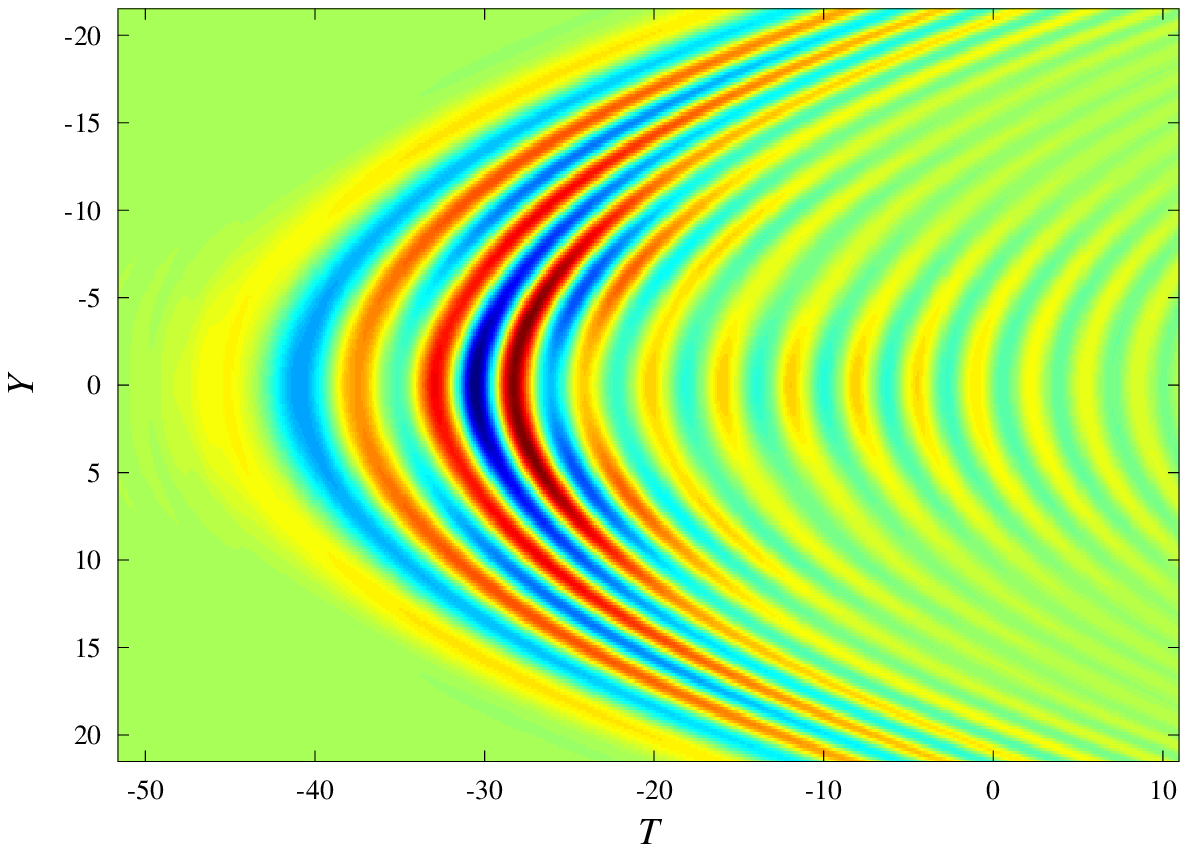}}
\subfigure[ ]{\includegraphics[width=7cm]{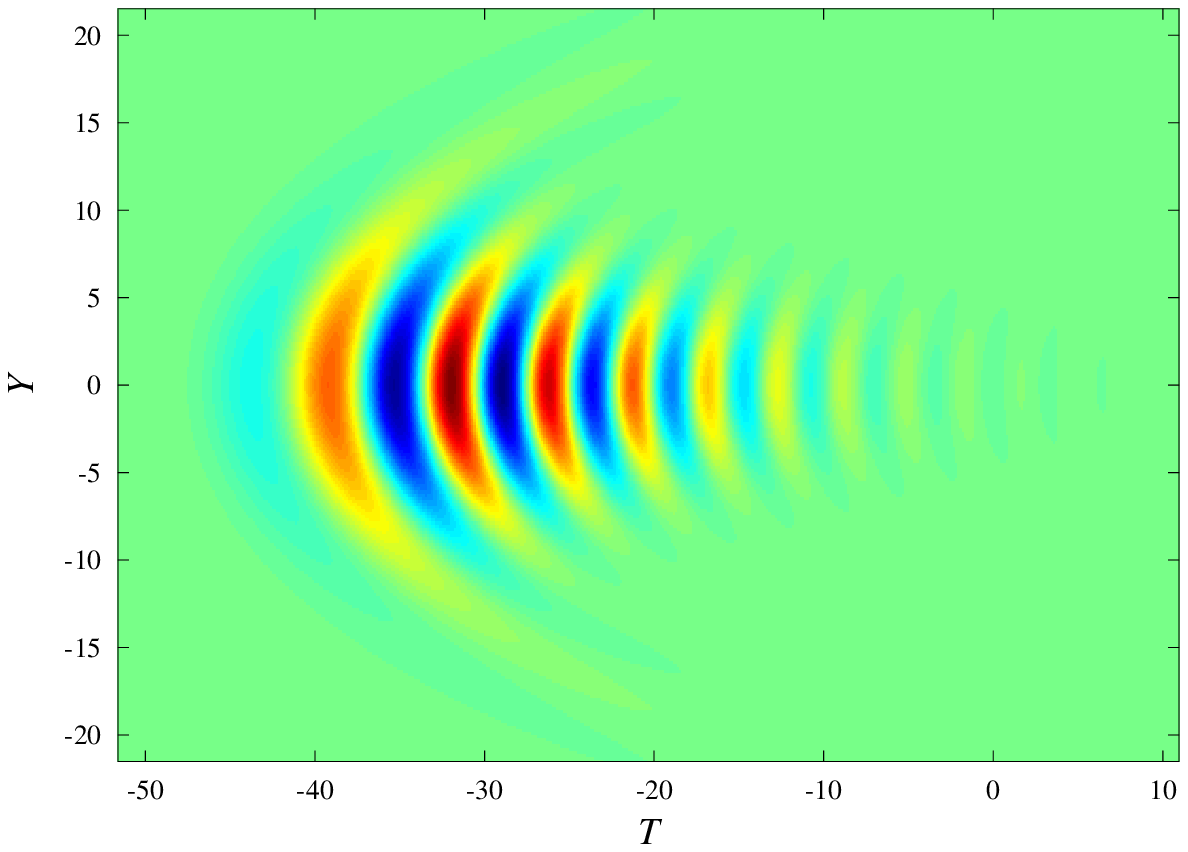}}
\caption{\cor{(Color online) Nonlinear diffraction of a FCP. a) Initial data ($Z=0$). b) An intermediary stage, with a typical crescent shape 
($Z=0.9594$). c) Nonlinear diffraction at $Z=3.984$. d) Linear diffraction at the same propagation distance for comparison.
Input data is expression (\ref{data}) with
$p=4.0825$, 
$q =  2.8868$, 
$\omega=1$, $A=4.8990$ 
(a, b, c), and $A=10^{-7}$ (d).}}
\label{diff}
\end{center}\end{figure}

\cor{Setting $\sigma_1=\sigma_2=-1$ as pertains to the Maxwell-Bloch system, we observe nonlinear diffraction (Fig. \ref{diff}). 
The nonlinear effect strongly increases the diffraction (compare Figs. \ref{diff}c and d). 
The conjugated effect of temporal self-compression and diffraction may lead to an intermediary stage, 
in which the pulse is very well localized temporally, and strongly widens spatially. This leads to a 
characteristic crescent shape (Fig. \ref{diff}b).
For the computations 
we used the parameters
$p=4.0825$, 
$q =  2.8868$, 
$A=4.8990$, 
and $\omega=1$.}

\cor{
The  CGKP equation (\ref{mkp}) can be generalized using the
expressions (\ref{coefa}) and (\ref{coefb}) of the coefficients. For a medium with anomalous dispersion and focusing nonlinearity, 
$A,\,B<0$ and $\sigma_1=\sigma_2=+1$. In this case spatiotemporal self-focusing occurs.}
\begin{figure}\begin{center}
\subfigure[ ]{
\includegraphics[width=8cm]{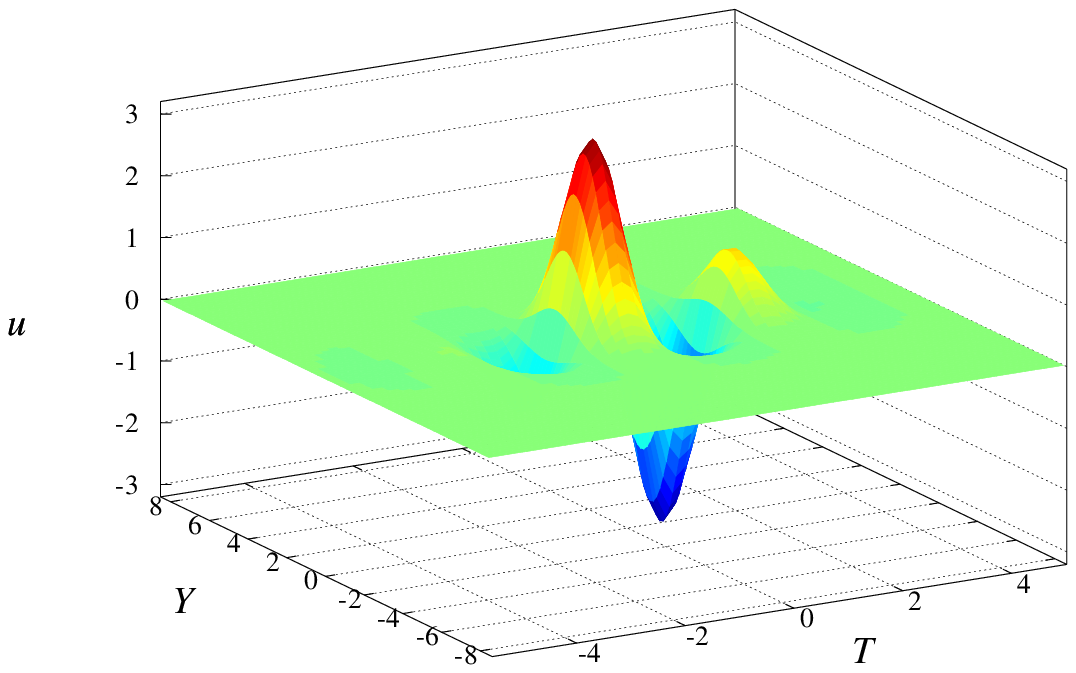}}
\subfigure[ ]{
\includegraphics[width=8cm]{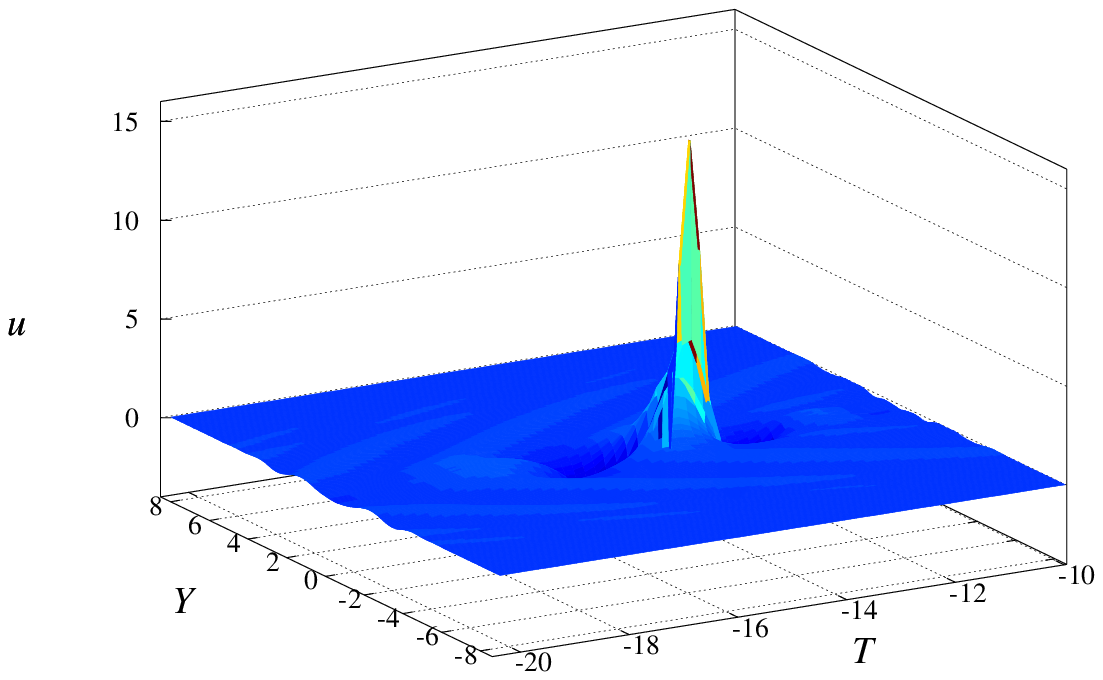}}
\caption{(Color online) The collapsing pulse. a) Input; b) Last computed point. 
Input data is expression (\ref{data}) with
$\omega=-2.1909$, $p=1.8257$, $q=1.414$, and 
amplitude $A=3.65$ \cor{(curve $b$ in Fig. \ref{maxu} below)}.}
\label{fig_i}
\end{center}\end{figure}
\begin{figure}\begin{center}
\includegraphics[width=8cm]{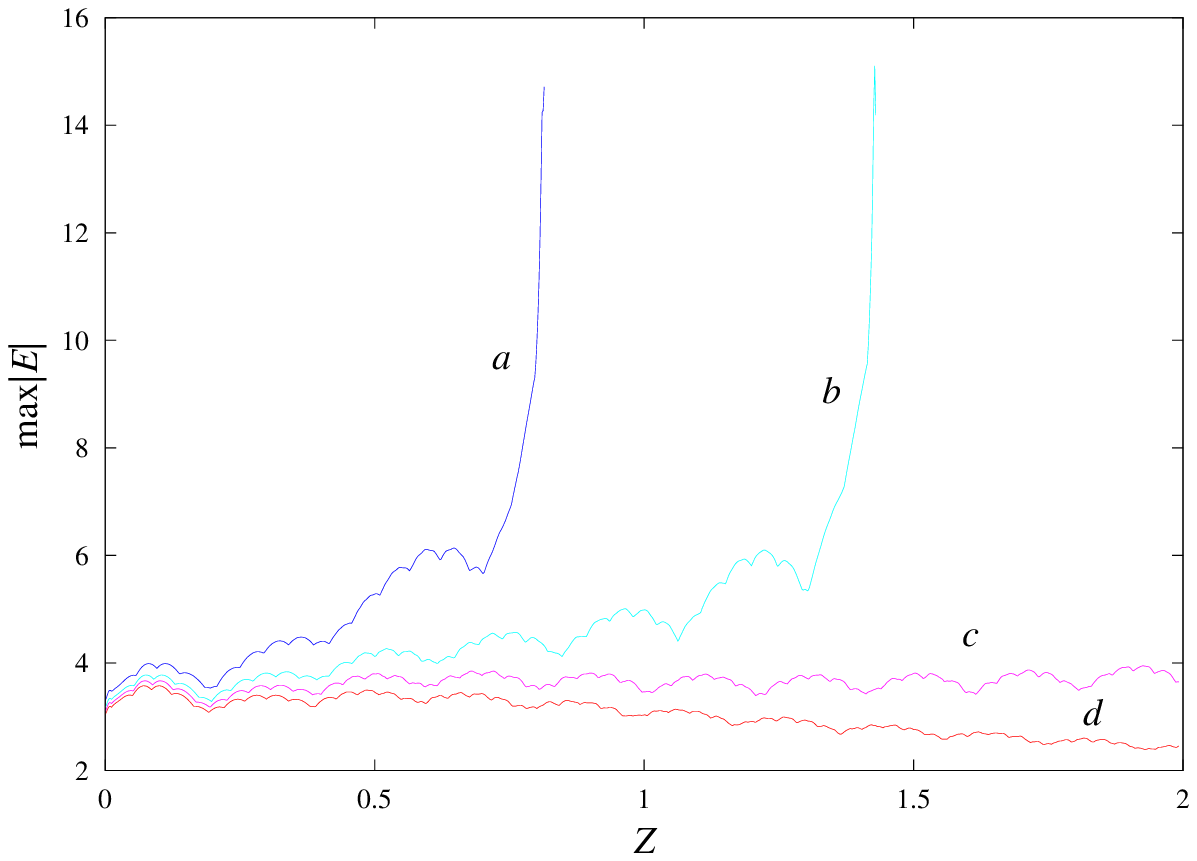}
\caption{(Color online) Evolution of the maximal value of the electric field for a few values of the initial amplitude. The 
collapse occurs above some amplitude threshold. Input data is expression (\ref{data}) with
$\omega=-2.1909$, $p=1.8257$, $q=1.414$, and several values of
the amplitude $A$, namely $A=3.80$ ($a$), 3.65 ($b$), 3.57 ($c$)  3.50 ($d$).
}
\label{maxu}
\end{center}\end{figure}

\cor{
Numerical resolution has been performed for $\omega=-2.1909$, $p=1.8257$, $q=1.414$, 
 }
and for several values of
the initial amplitude $A$, especially $A=3.80$, 3.65, 3.57, and 3.50.
Clear numerical evidence for collapse is found, see Figs. \ref{fig_i} and \ref{maxu}.
Collapse occurs for the two highest values of the input spatiotemporal field amplitudes $A$ (curves $a$ and $b$ in Fig. 2), and not for the two lowest  ones (curves $c$ and $d$ in Fig. 2).
Hence the occurence of some input amplitude threshold $A_{th}$ is evidenced, and for the considered pulse shape, frequency, length and
width,  we get the numerical estimation  $3.57<A_{th}<3.65$.

On the other hand, a rigorous mathematical analysis  of the CGKP equation (\ref{mkp}) has proved that wave collapse do occurs (for a comprehensive review of wave collapse in optics and plasma waves, see Ref. \cite{Luc}); also, for more details concerning CGKP equation see Refs. \cite{tur85a}-\cite{liu} and the analysis performed in the next Section.

\section{Calculation of the collapse threshold}

\cor{
In the following let us consider the generalized KP (GKP) equation in its normalized form 
\begin{equation}
\left(u_Z+u^pu_T+u_{TTT}\right)_T=u_{YY}. \label{mkpn}
\end{equation}
The focusing  CGKP equation (\ref{mkp}) with $\sigma_1=\sigma_2=+1$ is obtained for the particular case $p=2$.
}
\cor{The GKP equation possesses the conserved Hamiltonian
\begin{equation}
 H=\int\int\left[\frac12 v_Y^2+\frac12u_T^2-\frac1{(p+1)(p+2)}u^{p+2}\right]dTdY
\end{equation}
in which $v=\int^T u$. Note that the existence of the conserved Hamiltonian $H$  assumes  that both $u$ and $v_Y$ vanish at $+\infty$ and 
$-\infty$, which implies that 
$$\int_{-\infty}^{+\infty}udT=0,$$
i.e., that the mean value of the electric field is zero.}
\begin{figure}\begin{center}
\includegraphics[width=8cm]{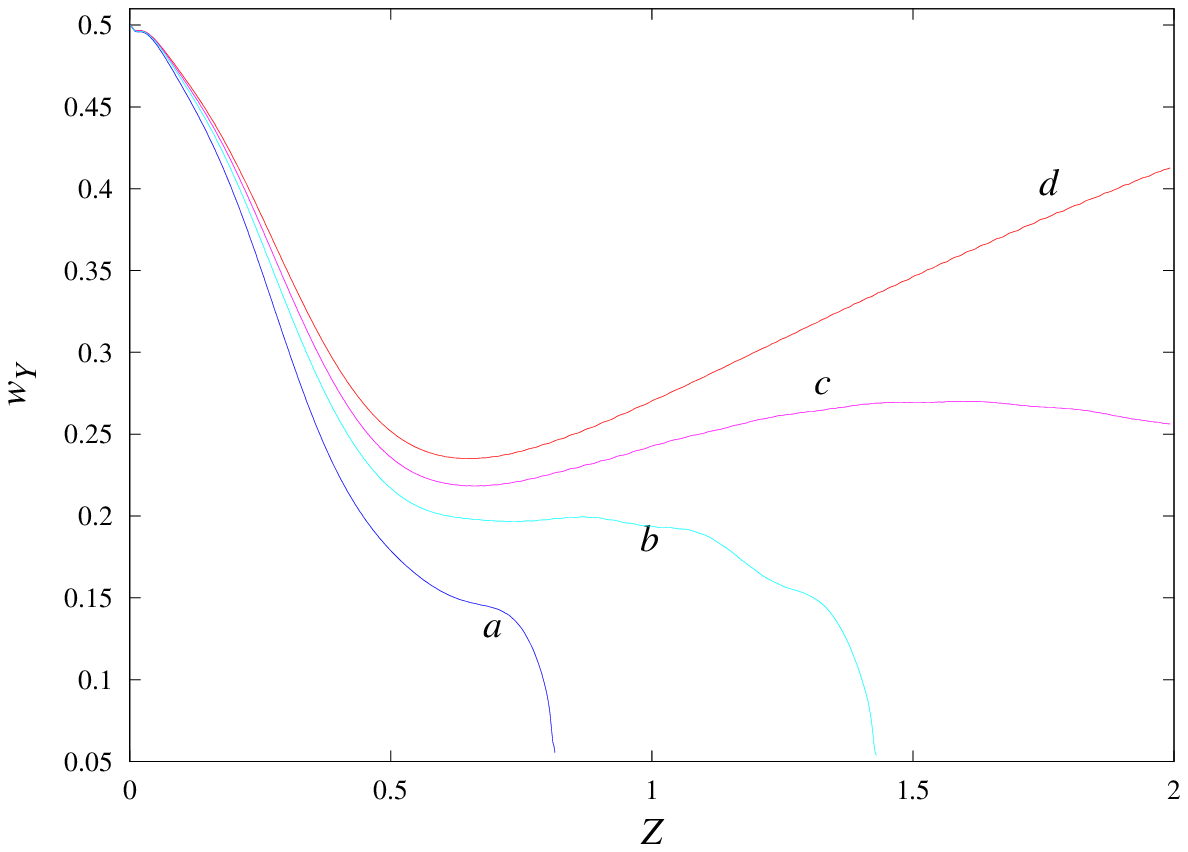}
\caption{(Color online) Evolution of the width $w_Y$ of the pulse, for the same initial amplitudes as in Fig. \ref{maxu}.
Self-focusing occurs initially for all considered data.}
\label{momy}
\end{center}\end{figure}

Using  a virial theorem, it has been proved that collapse occurs for $p>4$ and $H<0$ \cite{tur85a,wang94}.
However, these conditions are sufficient but not necessary.
Then, it was shown that the solitary wave is unstable for $p>4/3$  \cite{wang94,bou96b}, and a proof of collapse
by using a virial theorem was given in Ref. \cite{liu} for $4/3<p<4$, which includes the particular case $p=2$, which is relevant for the study of few-cycle optical pulses.
The assumptions of the theorem  involve the functional
\begin{equation}
Q(u)=\int\int\left[v_Y^2+u_T^2-\frac{3p}{2(p+1)(p+2)}u^{p+2}\right]dTdY.
\end{equation}
Together with regularity conditions and conditions involving the solitary wave solution, the main assumption of the 
blow-up theorem  is that the functional $Q(u)<0$,
in which $u=u(T,Y,Z=0)$ is the initial data.
For the expression of $u$ given by Eq. (\ref{data}), the functional $Q(u)$ can be exactly computed using standard methods, and we get
\begin{eqnarray}
Q(u)=\frac{A^2\pi}{128pq}&\Biggl\{&A^2p^2q^2\left(4e^{\frac{-\omega^2p^2}4}-3-e^{-\omega^2p^2}\right)
+32q^2\left(\omega^2p^2+1-e^{\frac{-\omega^2p^2}2}\right)\nonumber \\
&&+8p^3\sqrt{2\pi}\,e^{\frac{-\omega^2p^2}2}\int_{-\infty}^{+\infty}
 \mbox{Im}\left[\mbox{erf}\left(\frac{T}p+\frac{i\omega p}2\right)\right]dT\Biggr\},\label{qu}
\end{eqnarray}
where Im is the imaginary part and erf is the error function.

The factor which multiplies $A^2$ in the wide bracket in Eq. (\ref{qu}) is always negative, and hence 
$Q(u)$ is negative for $A$ larger than some threshold value $\tilde A_{th}$, 
with

\begin{eqnarray}
\tilde A_{th}=\sqrt{\frac{
32q^2\left(\omega^2p^2+1-e^{\frac{-\omega^2p^2}2}\right)
+8p^3\sqrt{2\pi}\,e^{ \frac{-\omega^2p^2}2}\int_{-\infty}^{+\infty}
 \mbox{Im}\left[\mbox{erf}\left(\frac{T}p+\frac{i\omega p}2\right)\right]dT}
{p^2q^2\left(3+e^{-\omega^2p^2}-4e^{\frac{-\omega^2p^2}4}\right)}
}
,\label{ath}
\end{eqnarray}

For the specific values of  parameters used in our numerical computations, we find that $\tilde A_{th}=7.567$.
 The threshold $A_{th}\simeq 3.6$ found numerically
is about half of this value $\tilde A_{th}$, found using the assumptions of the above mentioned virial theorem, which is consistent with the well-known fact that the conditions of the theorem of Ref. \cite{liu}, especially $Q(u)<0$,
yield  only a sufficient condition.
The quantity $\tilde A_{th}$ will be referred to below as the `analytical threshold for collapse' for the sake of simplicity. 

However, it is likely that this condition is optimal for an input which has a shape adapted to the collapse process.
On the basis of this idea and of the numerical results,
the discrepancy can be physically  interpreted as follows.
Figure \ref{momy} presents the evolution of the width $w_Y$ of the pulse against propagation distance $Z$, for a few initial values of input amplitude $A$ (see Eq. (\ref{data})). 
The width $w_Y$ is numerically computed according to:
$$w_Y^2={\int u^4Y^2dTdY \over \int u^4dTdY}.$$
It is seen that self-focusing occurs in every case presented in Fig. 4. However, for initial amplitude
below threshold $A_{th}$, the self-focusing stops after a while and the  collapse is inhibited. This
feature is due to the dispersion (both linear and nonlinear), which tends to increase the temporal length of the pulse 
at the same time as it self-focuses. Below the threshold, the dispersion dominates and collapse is prevented,
while above the threshold, self-focusing dominates and collapse occurs.
It is worthy to mention that the arrest of collapse due to dispersion was already mentioned  in Ref. \cite{igor2000}.

This observation may justify qualitatively the discrepancy between the analytic and numerical thresholds for collapse found above.
Between the two values for threshold ($A_{th}\simeq 3.6\lesssim A \lesssim7.6\simeq \tilde A_{th} $), at the beginning of the process, the amplitude is not
properly speaking  sufficient to initiate collapse but, due to the shape of the pulse, a nonlinear lens effect induces a transverse 
self-focusing of the pulse, which increases the maximal pulse amplitude. At the same time, dispersion (linear and nonlinear) occurs,
which tends to decrease the amplitude. If dispersion dominates, the growth of the amplitude stops and collapse does not occur. If, on the contrary, self-focusing dominates, the peak amplitude reaches a value which is sufficient to induce the collapse as such.
Notice that in Fig. \ref{maxu}, the collapsing curves show two distinct parts, the first one (oscillating) corresponds 
rather to self-focusing and the second one corresponds rather  to collapse \textit{stricto sensu}. The value of the amplitude at the 
boundary between the two regions is close to the 
threshold value for collapse ($\tilde A_{th}\simeq 7.6$) found from the mathematical condition $Q(u)<0$, see Eqs. (\ref{qu}) and (\ref{ath}).

\section{Spectral broadening}

\begin{figure}\begin{center}
\includegraphics[width=8cm]{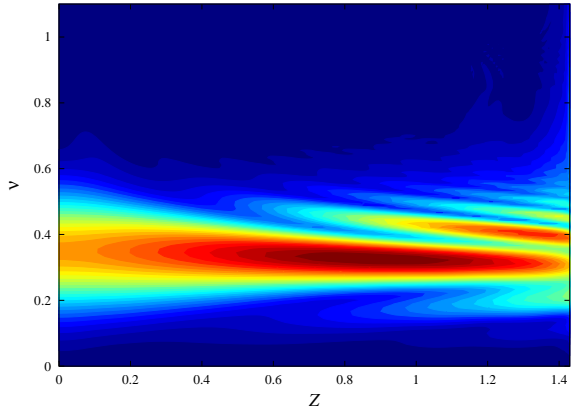}
\caption{(Color online) Evolution of the spectrum integrated over $Y$ during collapse. The input data correspond to the case $b$ in Fig. \ref{maxu}.}
\label{spec_g}
\end{center}\end{figure}
\cor{
\begin{figure}\begin{center}
\subfigure[ ]{\includegraphics[width=9cm]{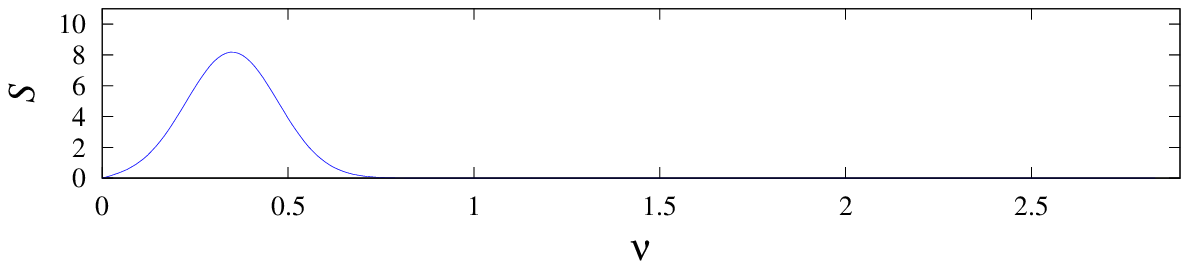}}
\subfigure[ ]{\includegraphics[width=9cm]{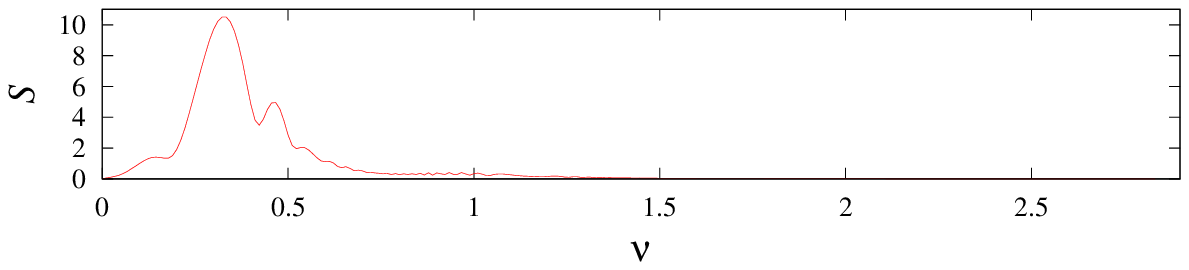}}
\subfigure[ ]{\includegraphics[width=9cm]{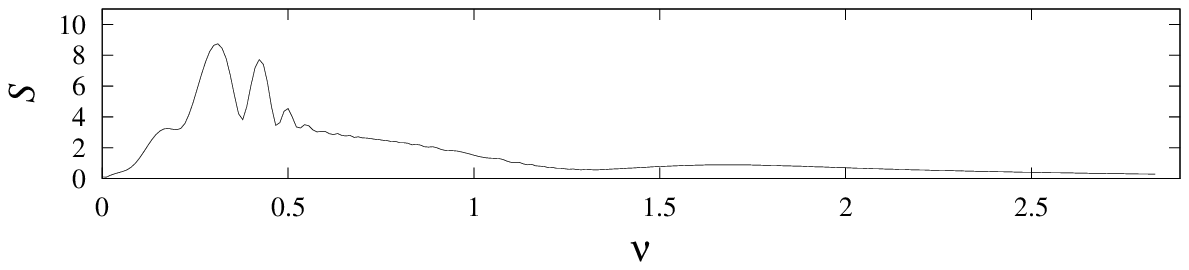}}
\caption{\cor{(Color online) Evolution of the spectrum $S$ (integrated over $Y$) during collapse. a) Initial data ($Z=0$). b) At the limit between the `self-focusing' and 'collapse \textit{stricto sensu}' stages, $Z=0.623$.
c) At the last computed time $Z=0.81445$. The input data correspond to the case $a$ in Fig. \ref{maxu}.}}
\label{specl_g}
\end{center}\end{figure}
}

The evolution of the spectrum (integrated over $Y$) is shown in \cor{ Figs. \ref{spec_g} and \ref{specl_g}}. 
At the  beginning of the evolution process, the spectrum tends to be narrowed. Then the spectrum presents the 
oscilatory structure typical of the modulation instability \cor{(Fig. \ref{specl_g}b)}. However it is strongly asymmetric, in contrast with the 
case of the `long' pulses described within the SVEA. Finally, a strong broadening of the spectrum is seen as the collapse itself occurs \cor{(Fig. \ref{specl_g}c)}.

\section{Conclusions}

In conclusion, we have introduced a model beyond the
slowly varying envelope approximation of the comonly used nonlinear
Schr\"{o}dinger-type evolution equations, for describing the
propagation of (2+1)-dimensional spatiotemporal ultrashort
optical solitons in Kerr (cubic) nonlinear media. Our approach
is based on the Maxwell-Bloch equations for an ensemble of
two level atoms and on the multiscale approach, and
as a result of using the powerful reductive perturbation method \cite{tutorial},
a generic cubic generalized Kadomtsev-Petviashvili partial differential evolution equation was put
forward. \cor{
Nonlinearly enhanced  diffraction accompanying temporal self-compression is observed.
In the case of anomalous dispersion and focusing nonlinearity,
collapse is evidenced.}
The collapse threshold for the propagation of ultrashort spatiotemporal pulses described
by the cubic generalized Kadomtsev-Petviashvili equation was calculated 
numerically, and compared to the analytical results  
drawn from a theorem based on a virial method, which proves that collapse occurs.
The discrepancy between the value of the collapse treshold obtained numerically and the corresponding analytical threshold for collapse 
is qualitatively explained. Moreover, the evolution of the spectrum (integrated over the transverse, spatial coordinate) 
is also given and a strongly asymmetric spectral broadening of ultrashort
pulses during collapse is also put forward, in contrast to the case of long spatiotemporal pulses 
described within the slowly varying envelope approximation. 
The present study is restricted to (2+1) dimensions, however, it can be extended to the (3+1) dimensions \cite{LB} by incorporating into the generic model a second transverse (spatial coordinate).

\section{Acknowledgment}

One of the authors (D.M.) was supported in part by the Romanian Ministry of Education and Research through Grant No. IDEI-497/2009.

\end{document}